# CASPIANET++: A Multidimensional Channel-Spatial Asymmetric Attention Network with Noisy Student Curriculum Learning Paradigm for Brain Tumor Segmentation

Andrea Liew[a], Chun Cheng Lee[b], Boon Leong Lan[a,c], Maxine Tan[a,d*]

[a]Electrical and Computer Systems Engineering Discipline, School of Engineering, Monash University Malaysia, Bandar Sunway 47500, Malaysia

[b]Radiology Department, Sunway Medical Centre, Bandar Sunway 47500, Malaysia

[c]Advanced Engineering Platform, School of Engineering, Monash University Malaysia, Bandar Sunway 47500, Malaysia

[d]School of Electrical and Computer Engineering, The University of Oklahoma, Norman, OK 73019, USA

*maxine.tan@monash.edu

## Abstract

Convolutional neural networks (CNNs) have been used quite successfully for semantic segmentation of brain tumors. However, current CNNs and attention mechanisms are stochastic in nature and neglect the morphological indicators used by radiologists to manually annotate regions of interest. In this paper, we introduce a channel and spatial wise asymmetric attention (CASPIAN) by leveraging the inherent structure of tumors to detect regions of saliency. To demonstrate the efficacy of our proposed layer, we integrate this into a well-established convolutional neural network (CNN) architecture to achieve higher Dice scores, with less GPU resources. Also, we investigate the inclusion of auxiliary multiscale and multiplanar attention branches to increase the spatial context crucial in semantic segmentation tasks. The resulting architecture is the new CASPIANET++, which achieves Dice Scores of 91.19% whole tumor, 87.6% for tumor core and 81.03% for enhancing tumor. Furthermore, driven by the scarcity of brain tumor data, we investigate the Noisy Student method for segmentation tasks. Our new Noisy Student Curriculum Learning paradigm, which infuses noise incrementally to increase the complexity of the training images exposed to the network, further boosts the enhancing tumor region to 81.53%. Additional validation performed on the BraTS2020 data shows that the Noisy Student Curriculum Learning method works well without any additional training or finetuning.

## Introduction

Brain tumors are uncontrolled growth of cells in the brains, which affects essential bodily functions such as breathing. Treatment of brain tumors typically consists of irradiating the tumor bed with high doses of radiotherapy. To minimize the impact of radiation on healthy cells, accurate determination of tumor volumes are crucial [1]. The task of registering these tumor volumes is performed by manually outlining hundreds of MRI images per patient. This process presents its own set of challenges as tumor boundaries are easily mistaken for healthy tissue and could be missed due to poor contrast and small size. Despite aggressive treatment given to patients, the average survival rate is 2 years [2]. One factor for the poor prognosis is accuracy of tumor segmentation, as treatment planning depends on precise demarcation of tumor regions. This is further exacerbated by low interobserver agreement between radiologists in determining the tumor regions [3]. With an estimated 84,170 new primary brain tumor diagnoses in 2021, there is an urgent need to develop automatic segmentation algorithms for quick and accurate segmentation [4].

Today, segmentation algorithms have evolved from classical machine learning methods, which rely on hand crafted features, to deep learning models which have achieved far greater accuracies than the former. Deep learning architectures, specifically Convolutional Neural Networks (CNNs) have been used in many brain tumor segmentation algorithms [5][6][7]. An improvement over the CNN, Fully Convolutional Networks (FCN) are capable of generating dense predictions for segmentation tasks, allowing larger input images to be processed [8]. One such FCN is UNET, which has been developed for biomedical image segmentation tasks [9]. As GPU capacity increases, deeper and more complex architectures inspired by these models are being built. However, in the interest of efficiency and energy conservation, less complex models, which can provide similar accuracy without the additional computational overhead, are preferred.

One promising concept is the Attention mechanism, which was initially used in natural language processing [10], but has found its way into medical image segmentation tasks. Attention gates (AG) are embedded into the skipped connections of UNET to highlight regions of saliency by incorporating spatial information from coarser scales to drive the upsampling operation [11]. Similarly, Squeeze and Excitation (SE) layers was introduced to compute channel-wise importance of feature maps [12]. Both methods suggest spatial and channel information with higher information density are assigned a higher weightage. However, AG and SE layers rely on stochastic methods and global average pooling to calculate the spatial and channel importance. Pooling operations may extract regions with the highest intensities, which may not correlate with regions of the highest anomalies. This is particularly true for the brain region, where some normal structures of the brain can exhibit high contrast on the MRI.

In this article, we introduce a method based on a-priori information to address the issue of using pooling operations to determine the region of saliency. This a-priori information is based on the well-established notion that radiologists use to identify tumor regions known as bilateral symmetry [13][14]. This symmetry is prevalent in healthy brain structures, where the structure of the left and right hemisphere of the brain appear to be almost identical. Conversely, the presence of lesions and tumors distort the symmetry between the two hemispheres. Also, we do not discount the power of stochastic methods used in the AG and SE mechanisms to further enhance the accuracy of these regions. Therefore, we design a hybridized layer which uses a deterministic measure of asymmetry in the brain to encode a radiologist's insight into our attention layer. Morphologically, tumors are never uniformly shaped, so we introduce a multiplanar approach where asymmetry is evaluated on sagittal, coronal, and axial information. To increase the amount of spatial context crucial for semantic segmentation tasks, we introduce a multiscale approach via a gating mechanism to account for any spatial context which would have otherwise been lost.

In contrast with other deep learning applications for computer vision tasks, medical segmentation tasks present their own set of problems as there is only a handful of fully-annotated datasets available to train the deep learning models. In comparison, the publicly-available brain tumor segmentation dataset, BraTS 2018[15] contains 285 fully annotated images compared to ImageNET[16], which has 50,000 images. In low data regimes, it may be advantageous to explore the possibility of using unlabelled data to generate pseudo labels for self-training [17]. Self-training methods such as Noisy Student have proven itself in classification tasks [18]. The viability of this method was also shown to be successful in semantic segmentation tasks [19]. However, both experiments were conducted on generic, natural images with considerably larger datasets than the brain tumor dataset. Here, we investigate how to best leverage the Noisy Student method given the size constraints of our segmentation dataset. Central to the Noisy Student method is the addition of noise in the form of data augmentations, dropouts and stochastic depth [18]. However, training losses can spike when presented with difficult examples, which leads to instability. This condition is further exacerbated by uncertainty from pseudo labels and noise added into the training data. Hence, we propose a Noisy Student Curriculum Learning which incrementally adds noise to the samples.

The contributions of our paper are summarized as follows:

1. We introduce a novel **Ch**annel **Sp**atial **A**symmetric Atte**n**tion (CASPIAN) Layer based on a priori asymmetry information, where highly-dissimilar regions are given higher importance and propagated through the network.

2. We adopt a **multidimensional** approach integrating multiscale and multiplanar (i.e., axial, coronal and sagittal) information into the CASPIAN layer, resulting in the CASPIAN++ (**CASPIAN** + Multiscale + Multiplanar) layer, to increase spatial information within the network

3. We propose an improved deep neural network architecture, **CASPIANET++** which integrates multiscale and multiplanar spatial attention with channel attention for more accurate segmentation

4. We propose a curriculum learning paradigm called *Noisy Student Curriculum learning* for brain tumor segmentation tasks

5. We show that the CASPIANET++ architecture *with Noisy Student Curriculum learning* achieves state-of-the-art results on the BraTS 2018 validation dataset.

RELATED WORK

### Brain Tumor Segmentation

CNN based methods form the basis of many contemporary brain tumor segmentation algorithms. DeepMedic uses a dual pathway convolutional approach with three-dimensional (3D) fully-connected Conditional Random Fields (CRF) to achieve accurate segmentation [7]. However, the fully-connected layers at the output layer necessitate smaller patches with lesser spatial context during training. Architectures such as Fully Convolutional Networks (FCN) and encoder-decoder architectures allow larger input patches by up-sampling to restore predictions to the input dimension [8][9].

UNET continues to be a popular choice in biomedical image segmentation tasks and have dominated the BraTS competition leaderboards up to today [20] [21] [22]. The winning model of BraTS 2020 competition, aptly named No-New NET, was trained on vanilla 3D UNET through meticulous hyperparameter selection [23]. Another notable work, NVDLMED replaced UNET encoders with ResNet blocks to increase the depth of the network, and another VAE branch added regularization to the architecture [21].

General purpose CNN methods which have been used successfully in segmentation such as Mask R-CNN and SegNet have been adapted for brain tumor segmentation tasks. SegNet, which has a fixed 64 feature maps per layer throughout the encoder and decoder, and uses pooling indices at its skipped connections, is more lightweight in memory usage compared to UNET [24]. A hybridized version of SegNet and UNET, U-SegNet was introduced to combine the desirable features in each network [25]. Mask R-CNN with an attention gating module has also been used for one-shot localization and segmentation of brain tumors in [26]. One advantage of this scheme is that both the localization and segmentation is carried out in one model.

While most schemes are known to treat multimodal MRI data as a multichannel input, a new idea of training a UNET in pairs, known as Modality Pairing was introduced in BraTS 2020 [22]. Experimental results show that a pair consisting of T1 and T1ce, T2 and FLAIR work well together. This is a good compromise to schemes where each modality is trained on a separate branch of SegNet [27].

As with all tumor segmentation problems, the proportion of healthy tissue outnumbers abnormal ones – creating severe class imbalance. Consequently, neural networks would be predisposed to predict a tissue as healthy. A priority-based sampling strategy is utilized in training the 3D SegNet where difficult samples are assigned higher priorities [28]. Another strategy is to employ cascaded models to first segment the whole tumor, then zeroing in on finer regions in subsequent models [6] [29]. Despite its effectiveness, this technique calls for multiple models to be trained individually, which consumes time and GPU resources.

For additional volumetric context, 3D methods are favoured over their 2D counterparts [21][7]. Nevertheless, the performance gains from 3D methods come at the cost of high computational resources. To overcome this, DMFNet uses channel grouping with dilated convolutions in UNET to reduce computational complexity while still maintaining a high degree of accuracy [30].

### Attention mechanisms

The main purpose of attention blocks is to bias the network to regions which are densely informative. Attention gates rely on convolutions and downsampling from coarser scales to extract salient features [31]. Another popular and efficient approach - Squeeze and Excitation - uses global average pooling to determine channel-wise attention [12]. Here, spatial information in each channel is compressed into a single weight, so that significant areas may be ignored. However, this may be detrimental for brain tumor segmentation tasks, since tumorous areas may be inadvertently left out, in favour of higher intensity regions. While the original SE layers focus on channel-wise attention, this was extended to include spatial and channel squeeze and excitation in [32]. The proposed Attention UNET architecture added AG to skip connections in the expansive path, neglecting to incorporate attention to the contracting path [31]. While these methods had used fully-connected and pooling layers to obtain averages of spatial or channel information, they may not reflect the exact tumorous regions.

Despite advancements in deep learning methods, techniques borrowed from classical machine learning work can be incorporated to further improve segmentation accuracy. Classical brain tumor segmentation methods mimic radiologists by utilizing the assumption of bilateral asymmetry of the brain to localize tumor regions [13][14]. One method uses fixed templates to determine asymmetry, which may not be able to account for structural variations in patients [13]. Another method requires an estimate of the mid-sagittal plane, which would introduce some uncertainty since there could be structural differences between patient scans [14]. In [33], the shortcomings described in the earlier methods are addressed by assuming a simple contralateral flip of feature maps. However, the method assumes features maps across channels carry similar information, negating any channel importance emphasized in the SE blocks.

Our method proposes to use the inherent asymmetry in tumorous organs [34][10][23] as attention maps. The aim of incorporating this asymmetric information in network layers is to highlight boundary regions which have proved difficult to segment due to lack of contextual information [35]. However, the authors in [33] highlighted that while the flip method fared better for whole tumor structures, it did not perform well on finer structures such as enhancing tumors. To address this, we incorporate multidimensional asymmetry, through a dual scale asymmetric attention scheme [36][37] calculated on three planar views to provide additional spatial context to the attention map.

### Semi supervised learning

Acquisition of fully-annotated medical images for supervised learning for segmentation tasks is limited and costly, compared to the availability of unlabelled images. To overcome this, transfer learning on encoders trained with out-of-domain ImageNet data was proposed for brain tumor segmentation [38]. However, as only three modalities were used in finetuning to conform with the RGB channels of a pretrained ResNet34 encoder, results did not favour a four-modality MRI dataset. Pseudo labelling was used to train labelled and unlabelled data together with a weighted unlabelled loss which increased with epochs, as the confidence of the pseudo labels improves [17]. In brain image and Covid-19 segmentation tasks, pseudo labels were augmented with MixUp with data from hidden layers in [39]. The Noisy Student method showed that adding noise during self-training improved classification tasks [18]. This study was extended to semantic segmentation, which showed that self-training helps accuracy where pre-training may not [19].

Much success has been demonstrated in self training for classification, particularly for large datasets. The application of self-training to semantic segmentation is briefly explored in [19]. However, for brain tumor segmentation, the Noisy Student method has yet to be explored. In line with the academic theme of teacher-student training, we incorporate the idea of curriculum learning to the teacher-student iterations. Curriculum learning advocates learning tasks with increasing difficulty to improve generalization of the network [40]. The difficulty of a sample could be determined by the disease severity classified by a physician [41]. However, there could be an inverse relationship between disease severity and the sample difficulty. Another easy-to-difficult strategy involves training the network in a coarse-to-fine fashion, to expose the network to large tumor regions before moving on to finer structures [35].

Here, we propose a simple yet effective way to increase the complexity of our data through the addition of noise. Our teacher model starts off with training on only labelled images from the dataset. Subsequently, we inject uncertainty by joint training with pseudo labels, and we progressively add more augmentations so that the models can learn more difficult scenarios.

## METHODS

Spatial and Channel Asymmetry share the same input $I$, which is the output of any arbitrary encoder or decoder of the CNN architecture. All the inputs in this section are feature maps with dimensions $HxWxDxC$, where $H, W, D, C$ correspond to the height, width, depth and number of channels, respectively.

### Spatial Asymmetric Attention Mask

Let feature maps obtained from a layer be $U$, consisting of feature vectors $\{u_{ijk}^1, u_{ijk}^2, \ldots \ldots u_{ijk}^C\}$ for a single point $i, j, k$ along the entire channel axis from $1$ to $C$, where $i = 1\ to\ H, j = 1\ to\ W, k = 1\ to\ D$. Similarly, $\widehat{U}$ is the horizontal-flip of $U$, with corresponding feature vectors $\{\hat{u}_{ijk}^1, \hat{u}_{ijk}^2, \ldots \ldots \hat{u}_{ijk}^C\}$.

$\alpha_{sp}(i,j,k)$, is a similarity measure, which quantifies how similar a particular pixel at location $i, j, k$ is relative to its horizontally flipped counterpart. It is defined as:

$$\alpha_{sp}(i,j,k) = \frac{\sum_{l=1}^{C} u_{ijk}^l \hat{u}_{ijk}^l}{\sqrt{\sum_{l=1}^{C} {u_{ijk}^l}^2} \sqrt{\sum_{l=1}^{C} {\hat{u}_{ijk}^l}^2}} \quad (1)$$

$\alpha_{sp}(i,j,k) \in \mathbb{R} \mid [-1,1]$. Values closer to -1 and 1 indicate high dissimilarity and high similarity, respectively.

Values of $\alpha_{sp}$ in the range [-1,1] are normalized to [0,1] to obtain $\theta_{sp}$, the Spatial Asymmetric Attention Mask (SAAM)

$$\theta_{sp} = \frac{1 - \alpha_{sp}}{2} \quad (2)$$

Dissimilar pixels indicate that the symmetry rule in the left and right regions have been violated and there is a high chance of tumor in that pixel. After normalization, regions of high dissimilarity have higher values and can be represented as a weighted mask, to give attention/importance to the tumorous regions.

### Channel Asymmetric Attention Mask

In [33], every channel was assumed to contain similar spatial information and carries equal importance. However, different tumor regions tend to exhibit better contrast in certain MRI modalities. A similar protocol is followed by experts in annotating the ground truth for the BraTS challenge [15]. Therefore, to account for the multimodal nature of the images, we extend the spatial attention method in [19] and propose to weight each channel by a scalar value, termed channel-wise asymmetric attention.

Let feature maps obtained from a layer be $M$, consisting of feature vectors $\{m_{1,1,1}^p, m_{1,1,2}^p, \ldots \ldots m_{H,W,D}^p\}$ for every point in the feature map at a single channel axis $p$, where $p = 1$ to $C$. Similarly, $\hat{M}$ is the horizontal-flip of $M$, with corresponding feature vectors $\{\hat{m}_{1,1,1}^p, \hat{m}_{1,1,2}^p, \ldots \ldots \hat{m}_{H,W,D}^p\}$.

$\alpha_{ch}(p)$, is a similarity measure, which quantifies how similar a particular channel $p$ is relative to its horizontally flipped counterpart. It is defined as:

$$\alpha_{ch}(p) = \frac{\sum_{ijk} m_{ijk}^p \hat{m}_{ijk}^p}{\sqrt{\sum_{ijk} {m_{ijk}^p}^2} \sqrt{\sum_{ijk} {\hat{m}_{ijk}^p}^2}} \quad (3)$$

$\alpha_{ch}(p) \in \mathbb{R} \mid [-1,1]$. Values closer to -1 and 1 indicate high dissimilarity and high similarity, respectively.

Values of $\alpha_{ch}$ in the range [-1,1] are normalized to [0,1] to obtain $\theta_{ch}$, the Channel Asymmetric Attention Mask (CAAM) as follows:

$$\theta_{ch} = \frac{1 - \alpha_{ch}}{2} \quad (4)$$

The resulting $\theta_{ch}$ is a scalar value denoting how dissimilar the feature map for each channel is, indicating which channel may hold more salient information.

### Channel and Spatial Asymmetric Attention (CASPIAN) Layer

In the literature, brain tumor segmentation methods rely solely on the bilateral asymmetry on the mid-sagittal plane regions to pinpoint the locality of tumors. This necessitates that the mid-sagittal plane be as accurate as possible. However, in our method we apply a geometric flip to the input feature maps to calculate its asymmetry mask. The geometric flip is merely a simplification to determine the bilateral asymmetry in lieu of calculating the mid-sagittal plane. The oversimplification in the left-right flip may however result in some unwanted regions of asymmetry. To overcome this, in the final step we multiply the asymmetry mask with the feature maps, to ensure that the high intensity from the tumor regions amplify the regions of asymmetry while the low intensity non-tumor regions attenuate the unwanted regions.

The similarity masks from equations (2) and (4) are scaled linearly from 0 to 1. The resulting attention mask alone, when multiplied by its input, was insufficient to amplify the salient regions. Therefore an excitation block (depicted in Figure 1), which consists of a fully connected layer, $FC$, a non-linear ReLU block, $R$, followed by another fully connected layer, $FC$, and sigmoid layer, $\sigma$, is first applied to the mask. While the original excitation block [12] had proposed a bottleneck layer for computational efficiency, we did not utilize a bottleneck layer to preserve the structural integrity of the asymmetric attention mask. The resulting asymmetric attention mask with excitation for spatial, $S_{spatial\ attn}$ and channel, $S_{channel\ attn}$ are obtained as follows:

$$S_{spatial\ attn} = \sigma(FC_2(R(FC_1(\theta_{sp}))) \quad (5)$$

$$S_{channel\ attn} = \sigma(FC_2(R(FC_1(\theta_{ch}))) \quad (6)$$

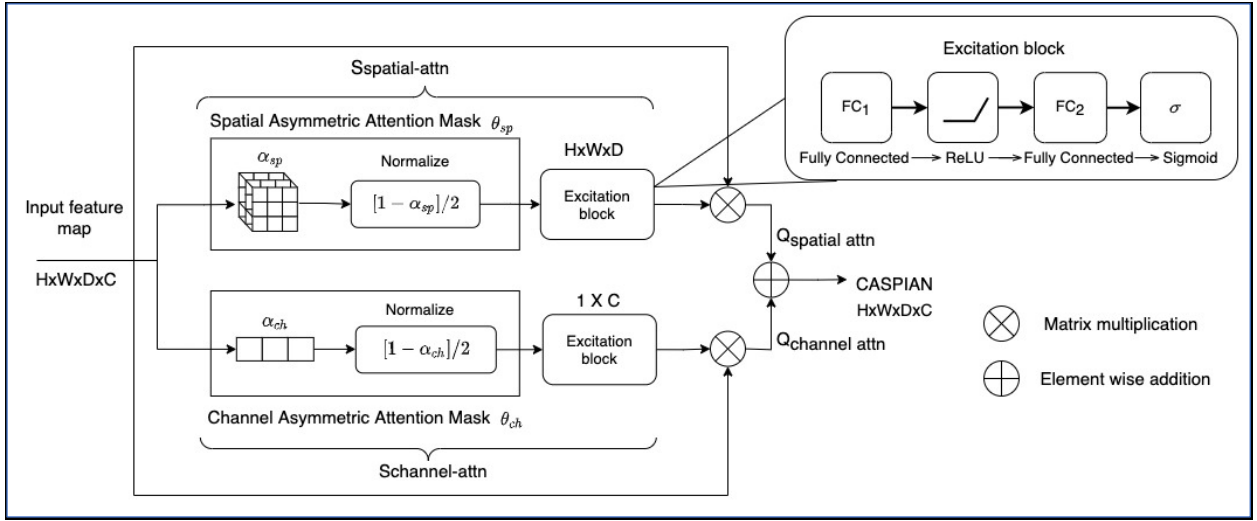

*Figure 1: The proposed channel and spatial asymmetric attention with excitation block (CASPIAN), where the input feature map is the output of any arbitrary encoder or decoder of the CNN architecture*

The spatial attention map $Q_{spatial\ attn}$ and the channel attention map $Q_{channel\ attn}$ are obtained from element-wise multiplication of the input $I$ :

$$Q_{spatial\ attn} = S_{spatial\ attn} \cdot I \quad (7)$$

$$Q_{channel\ attn} = S_{channel\ attn} \cdot I \quad (8)$$

The $CASPIAN$ block places equal importance to both channel and spatial attention, hence both maps are aggregated through element wise addition (as depicted in the final addition block in Figure 1):

$$CASPIAN = Q_{spatial\ attn} + Q_{channel\ attn} \quad (9)$$

Multiplanar and Multiscale approach in CASPIAN layers

Tumors are non-homogenous structures which have dissimilar boundaries when viewed from different planes (coronal, axial and sagittal) of the MRI data. This suggests that a similarity mask evaluated on a single view may not be adequate in capturing information for the entire volume of the tumor. We therefore introduce a Multiplanar Spatial Asymmetric Attention block as depicted in Figure 2.

From equation (5), we compute $S_{spatial\ attn,plane}$, the spatial asymmetry mask with excitation between the feature map and its geometrical flip along a plane. $S_{spatial\ attn,plane}$ is evaluated for *sagittal (plane=1), coronal (plane=2),* and *axial (plane=3) planes*.

$Q_{multiplanar}$, the Multiplanar Spatial Asymmetric Attention, is the sum of the element-wise multiplication of the $S_{spatial\ attn,plane}$ with the input image $I$ :

$$Q_{multiplanar} = \sum_{plane=1}^{3} S_{spatial\ attn, plane} \cdot I \qquad (10)$$

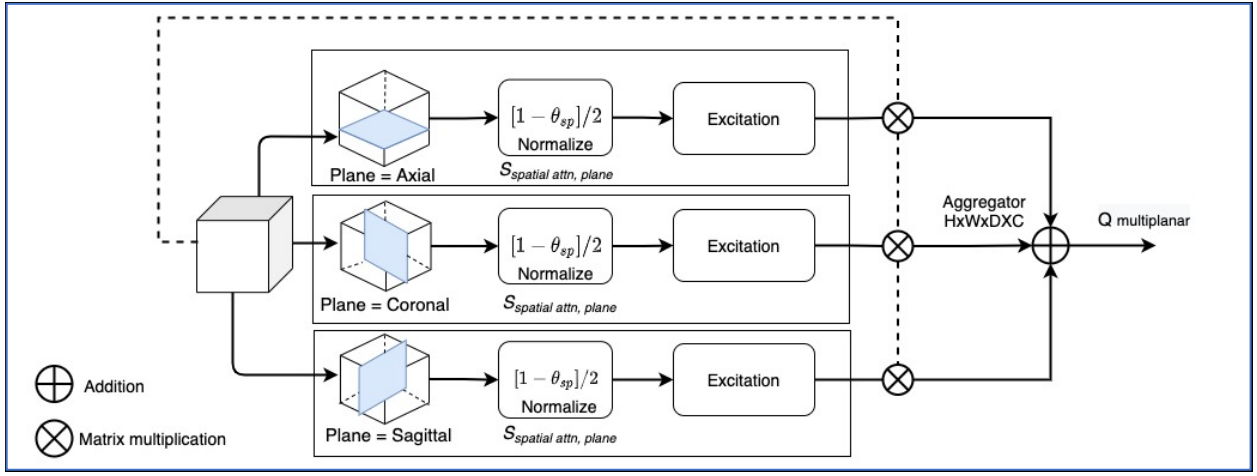

*Figure 2:* Multiplanar (axial, sagittal and coronal) Spatial Asymmetric Attention block

In delineating tumorous regions, the fusion of higher-level feature maps provides precise location and shape information while lower-level maps give structural context, such as texture and edges which are equally crucial in delineating tumorous regions. We therefore propose Multiscale Spatial Asymmetric Attention, $Q_{multiscale}$ to capture features across finer and coarser scales to delineate the tumors more accurately.

Consider a higher-level scale ($I_{S0}$), which typically is an input of the previous encoder block, and input at common scale ($I_{S1}$) in Figure 3. Here we introduce a gating mechanism to maximise the attention masks on both spaces. Since the most relevant information will be contained in $I_{S1}$, the spatial asymmetric mask $S_{spatial\ attn, s1}$ is given precedence.

To ensure precedence, $S_{spatial\ attn, s1}$ is multiplied with its image $I_{S1}$. To incorporate the information from $I_{S0}$ the signal of the mask from the common scale is inverted $(1 - S_{spatial\ attn, s1})$ to ensure any salient regions which are missed on the common scale can be captured by the higher scale. A maxpooling operation $P(I_{S0})$ is applied to $I_{S0}$, so that $I_{S0}$ is spatially equal to $I_{S1}$.

The Multiscale Spatial Asymmetric Attention $Q_{multiscale}$ is given by:

$$Q_{multiscale} = S_{spatial\ attn, s1} \cdot I_{S1} + (1 - S_{spatial\ attn, s1}) \cdot S_{spatial\ attn, P(I_{S0})} \cdot I_{S1} \qquad (11)$$

In the final block incorporating multiscale and multiplanar information into the CASPIAN layer we use an additive aggregator (see Figure 4), as our empirical studies show that the max-out layer does not have a positive effect on the validation results.

The resulting multiplanar and multiscale CASPIAN layer, CASPIAN++ is defined as follows:

$$\text{CASPIAN} + + = Q_{channel\ attn} + Q_{multiscale} + Q_{multiplanar} \qquad (12)$$

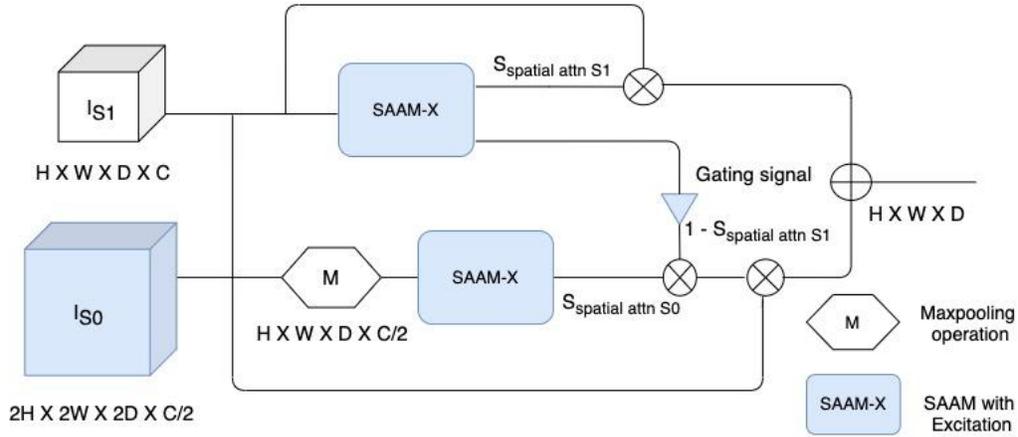

*Figure 3: The proposed multiscale approach with gating mechanism*

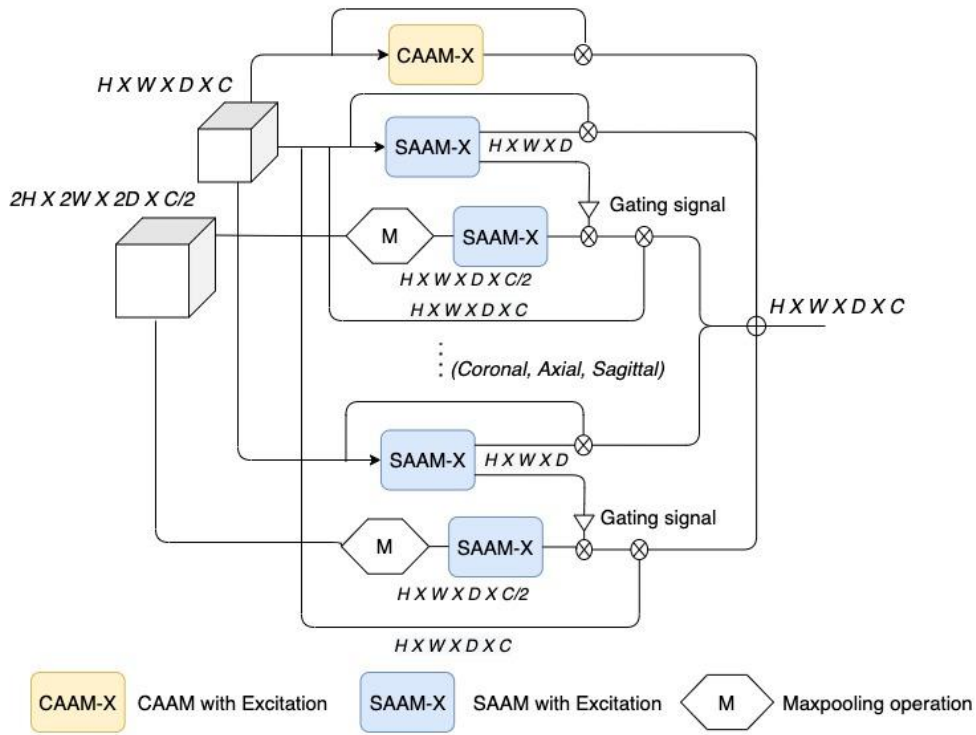

*Figure 4: Combining channel asymmetric attention into the multiscale and multiplane spatial asymmetric attention block*

CASPIANET++

The proposed CASPIAN++ layer can be integrated into any network to provide spatial and channel attention for semantic segmentation tasks. The network we chose is a lightweight variant of UNET with grouped convolutions as the base network [30]. Our architecture places CASPIAN/CASPIAN++ blocks at every path, except at the bottleneck (see Figure 5). This in effect recalculates asymmetry information for every feature map in the network, as spatial and contextual information may vary from layer to layer. A combination of CASPIAN and CASPIAN++ outputs are also added to skipped connections to filter irrelevant spatial information from earlier encoder layers for more accurate upsampling. No attention was added to the bottleneck, as spatial information has been substantially reduced from 128x128x128 to 8x8x8. Evaluating the spatial asymmetry on a layer with the least spatial information in the network would not be beneficial for segmentation tasks.

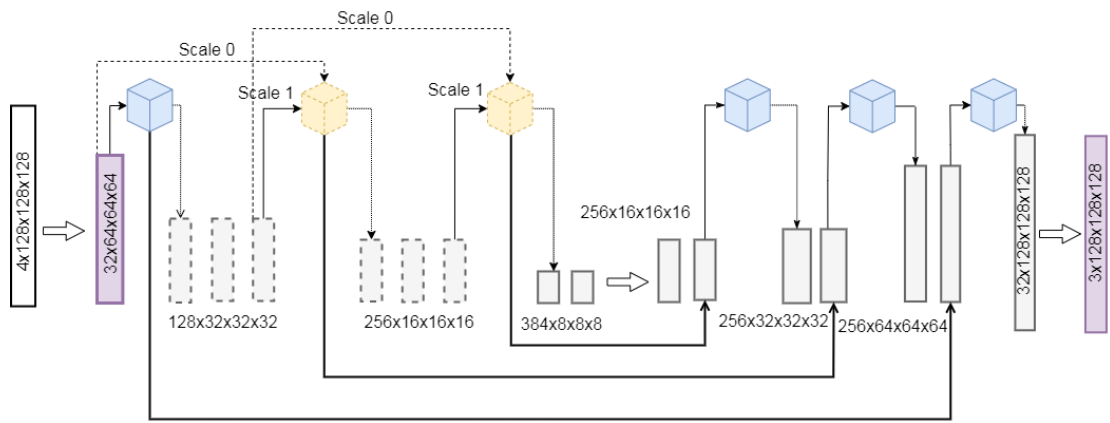

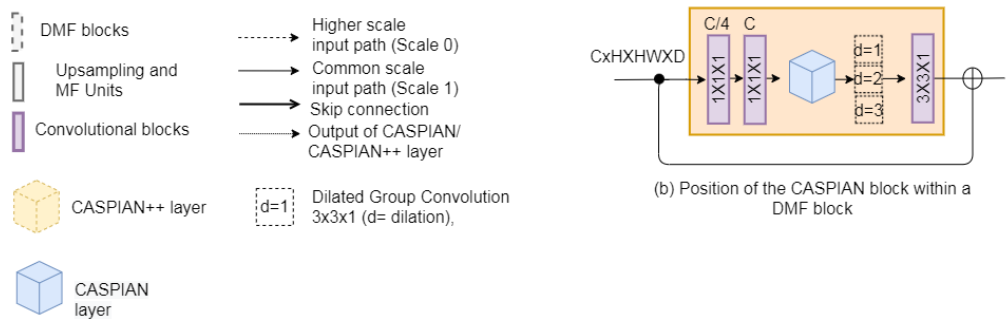

*Figure 5: The placement of CASPIAN and CASPIAN++ layers in the proposed CASPIANET++. 5(a) CASPIAN++ layers are added to the 2nd and 3rd encoder blocks. Notice that there are no CASPIAN layers at the bottleneck. 5(b) depicts the addition of the CASPIAN block within the DMF block*

CASPIAN++ was designed to include much more spatial information than the CASPIAN layer. Ideally, we would use CASPIAN++ layers in every block in the architecture. However, the multiscale approach on the first encoder block was memory intensive as it required another copy of the input image (128x128x128x4) to be stored in memory during training. Therefore, CASPIAN++ layers were added to certain layers only as a trade-off between hardware limitations and accuracy.

For maximum advantage, we incorporated CASPIAN++ layers into the second and third encoder blocks. These blocks were selected because they have the biggest receptive fields and would benefit most from the additional spatial information provided by CASPIAN++. Also, the increased spatial fields compensate for the conversion loss from spatial to contextual information. Within these blocks, CASPIAN layers are added within the DMF units to highlight salient features for the dilated convolutions (see Figure 5(b)). A similar scheme placed an attention layer after each dilated convolution [42]. However, for efficiency, we use one CASPIAN block before the dilations, to provide saliency to the dilated blocks. Hereon, the encoder-decoder implementation of CASPIAN/CASPIAN++ is known as CASPIANET++.

Region-based Loss Functions

The loss functions for training the network uses a region-based approach. Instead of optimizing for individual label classes (0, 1, 2, 4), the network treats the 3 regions $r$ as three separate binary problems. The three tumor regions are defined as a combination of labels as follows: whole tumor (WT) (labels 1,2 and 4), tumor core (TC) (labels 1 and 4) and enhancing tumor (ET) (label 4).

Consider the output of a region $r$, where $g_r$ represents the ground truth for the region and $p_r$ is the sigmoid of the predicted output of the region. The Binary cross entropy $\mathcal{L}_{bce}$ and Dice loss $\mathcal{L}_{dice}$ are defined as follows:

$$\mathcal{L}_{bce}(g_r, p_r) = -[g_r \log(p_r) + (1 - g_r) \log(1 - p_r)] \quad (13)$$

$$\mathcal{L}_{dice}(g_r, p_r) = 1 - \frac{2 * g_r * p_r}{g_r + p_r} \quad (14)$$

The training objective for the backpropagation step is to minimize the loss given by $\mathcal{L}_{total}$

$$\mathcal{L}_{total} = \sum_{r=1}^{3} \mathcal{L}_{bce}(g_r, p_r) + \mathcal{L}_{dice}(g_r, p_r) \quad (15)$$

where $r = 1,2,3$ for the respective regions of WT, TC and ET.

## Noisy Student Curriculum Learning

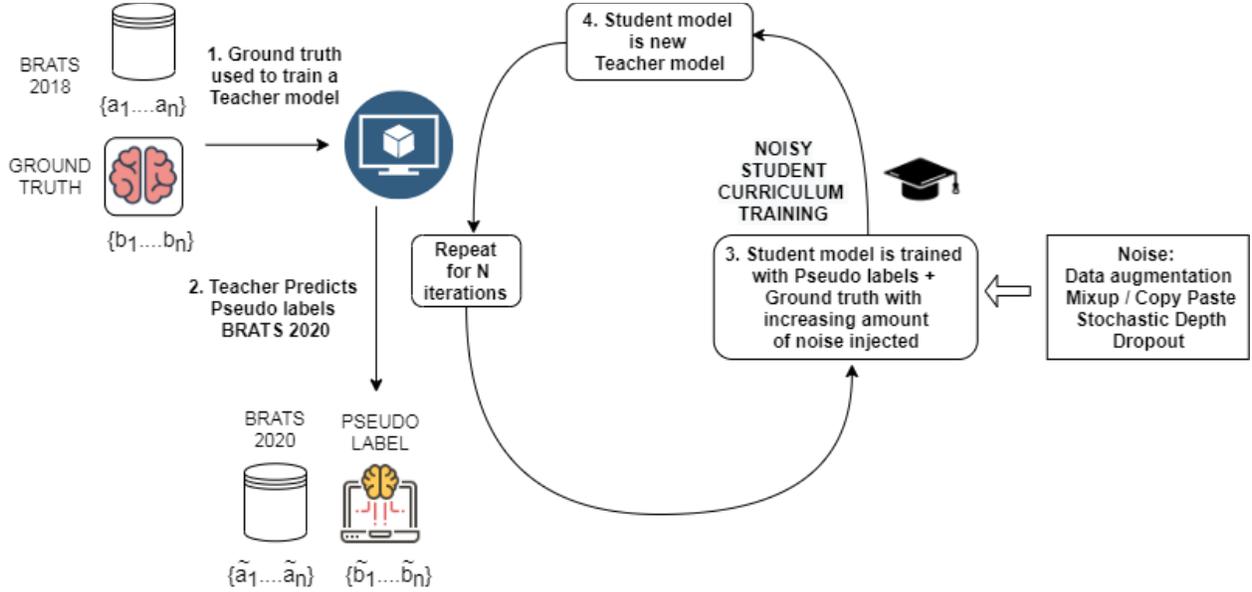

Figure 6 The Noisy Student Curriculum Learning workflow

Consider a set of images $\{(a_1, b_1), (a_2, b_2), (a_3, b_3), \ldots (a_n, b_n)\}$ where $(a_n, b_n)$ represents the image and labels of the $n^{th}$ patient in the dataset. The unlabelled set is denoted by $\{\tilde{a}_1, \tilde{a}_2 \ldots \tilde{a}_m\}$ where $\tilde{a}_m$ is the $m^{th}$ unlabelled image in the dataset.

Using the curriculum learning approach, noise is iteratively added to the student model so that it learns incrementally through an easy-to-hard approach; in this way, the student learns better as its parameters are first optimized on easy tasks followed by incrementally difficult tasks. All training data and pseudo labelled data are subjected to augmentation by centroid cropping, random rotations, intensity changes, flips and elastic deformations. This is to maintain consistency in training the teacher and student models.

In our method, two data augmentation methods were utilized to noise the pseudo labels, namely MixUp[43] and Copy-Paste [44]. Copy-Paste method was chosen as it demonstrates additive effects when used with other augmentation methods [44]. To increase difficulty in the pseudo labels, we experiment with the additive nature of Copy-Paste with MixUp.

Our proposed noisy student curriculum learning method (depicted in Figure 6) proceeds as follows:

1. Train a teacher model $\mathfrak{I}_t$ on the labeled dataset $\{(a_1,b_1),(a_2,b_2),(a_3,b_3),\ldots..(a_n,b_n)\}$ to minimize the loss function $\mathcal{L}_{total}$ in equation (15).

2. The teacher model is used to infer pseudo labels by $\{\tilde{b}_1,\tilde{b}_2\ldots..\tilde{b}_m\}$ for the unlabeled dataset $\{\tilde{a}_1,\tilde{a}_2\ldots..\tilde{a}_m\}$ as follows:

$$\tilde{b}_i = f(\tilde{a}_i,\mathfrak{I}_t), \forall\, i = 1\ldots..m \qquad (16)$$

3. Iteration of the Noisy student curriculum training starts here where $iter = 1,2,3$:

    3.1. Train a new student model $\mathfrak{I}_s^{iter}$ with noise $\eta_{iter}$ to minimize the loss function $\mathcal{L}_{total}$ on the labeled dataset $\{(a_1,b_1),\ldots..(a_n,b_n)\}$ and unlabeled dataset with their pseudo labels $\{(\tilde{a}_1,\tilde{b}_1),\ldots\ldots(\tilde{a}_m,\tilde{b}_m)\}$.

    The following are the three variations of $\eta_{iter}$ used
    1. MixUp
    2. Copy-Paste + Dropouts
    3. MixUp + Copy-Paste + Dropouts

    3.2. Use $\mathfrak{I}_s^{iter}$ as a new teacher model to infer new pseudo labels $\{\tilde{b}_1,\tilde{b}_2\ldots..\tilde{b}_m\}$ as per step (2).

    3.3. Repeat step 3.1 to train the next iteration of student model

### EXPERIMENTS AND RESULTS

#### Datasets

The proposed methods were validated on the BraTS 2018 dataset. BraTS 2018 dataset consists of 285 patients in the training set, with four available modalities, T1, T1c, T2, and FLAIR. For the Noisy Student Curriculum Learning, the BraTS 2020 validation and training data which are not part of the BraTS 2018 training set are used as the unlabelled dataset, amounting to 209 patients. This information is contained in the *name_mapping.csv* file provided by the organizer. The intensities of the four modalities are *z*-score normalized. We have also run some experiments on BraTS 2020 dataset to determine the generalization ability of our methods.

#### Implementation details

The architecture and layers proposed in this study are implemented in PyTorch. To demonstrate the performance of CASPIANET++, we had implemented these layers in the baseline model DMFNet [30]. The DMFNET is an enhancement of UNET, and was chosen because it is efficient and lightweight. A poly learning rate policy is implemented where the initial learning rate is 0.001 multiplied by a factor $[1-(e/E)^{0.9}]$ where $e$ is the current epoch and $E$, the total number of epochs. Synchronized batch normalization is used with a batch size of 4. Training is set to 300 epochs with an image crop size of 128x128x128. We maintain the following data augmentations throughout all experiments: (1) centroid cropping on WT regions, (2) random flipping, (3) random rotation, (4) random intensity and scaling, and (5) elastic deformations. For the optimizer, we utilize loss as the sum of region-based Dice Loss and BCE.

#### Ablation studies on components of the CASPIAN blocks

We perform experiments to demonstrate the improvement of each of the components towards the final CASPIANET++ block. We also include an implementation of the SE block (DMF+SE) to determine the contribution of channel-wise attention in multiclass semantic segmentation. We evaluate the performance of the model in terms of the Dice Scores and Hausdorff distance reported from the BraTS 2018 online evaluation platform.

From the results in Table I, the DMFNET+SE achieved dismal Dice scores, suggesting that a channel-only approach was insufficient to provide improvements to the pixelwise segmentation tasks. Compared to DMFNET+SE, the spatial approach in DMFNET+SAAM led to a 3.812% gain in the WT region. While a channel oriented approach did not do well on its own, adding the CAAM to the SAAM (DMF+SAAM+CAAM) further improved the gains in all three regions.

*Table I: Ablation studies on the effects of individual components of the CASPIAN blocks on Dice Score and Hausdorff95 - all models are trained for a total of 250 epochs for the BraTS 2018 Validation dataset*

|  | Dice Scores (%) | | | Hausdorff95 | | |
| --- | --- | --- | --- | --- | --- | --- |
| Model | ET | WT | TC | ET | WT | TC |
| DMFNET+SE | 78.709 | 87.044 | 83.112 | 3.233 | 24.240 | 8.009 |
| DMFNET+SAAM | 79.419 | 90.226 | 84.493 | 2.855 | 6.691 | 6.405 |
| DMFNET+SAAM+CAAM | 80.229 | 90.633 | 85.314 | 3.079 | 17.667 | 5.664 |
| CASPIANET | 80.338 | 90.826 | 85.616 | 2.895 | **4.717** | 5.570 |
| CASPIANET+MultiPlanar | 80.922 | 90.945 | 84.457 | 2.733 | 4.726 | 5.526 |
| CASPIANET+MultiScale | 80.749 | 90.986 | 86.703 | 3.159 | 6.394 | 5.589 |
| CASPIANET++ | **81.083** | **91.197** | **87.600** | **2.659** | 5.054 | **4.912** |

In the CASPIANET experiment, we introduced excitation to boost the values of the asymmetry masks which are closer to 1. Compared with the DMFNET+SAAM+CAAM, while the Dice Scores only showed marginal improvements of 0.1-0.3%, the Hausdorff95 distance showed marked improvements in all regions particularly in the WT region. This indicates that while the pixel accuracy did not change much, the predictions of the segmented regions are closer to the ground truth. Clinically, this would indicate that the shape of the segmentations more closely resemble that of the ground truth compared to the DMFNET+SAAM+CAAM.

Next, all multidimensional schemes are compared with a baseline, CASPIANET – to see how each variant performs individually and in CASPIANET++. First, we look at the multiplanar variant CASPIANET+MultiPlanar, which shows a 0.6% increase in the ET region. The extra symmetrical information, along all 3 plane views, falters in the TC region (-1.2%). On the other hand, the addition of multiscale information CASPIANET+MultiScale boosted the TC region (+1.1%). With the multiplane and multiscale combined, the CASPIANET++ essentially evaluates a multiscale CASPIAN block on three planar views, providing the network with a larger and more holistic spatial information of the tumor region. The results show a remarkable 1.98% improvement in the TC region from the CASPIANET architecture. When the multiplane and multiscale variants are ensembled as CASPIANET++, the overall scores achieved much better results compared to its individual components.

To illustrate the differences between the CASPIAN variants we present the predictions of sample Brats18_CBICA_AAL_1 in Figure 7(b) – (f). The prediction by DMFNET uses the model provided by the authors of DMFNET [30]. Compared to the ground truth in Figure 7(a), DMFNET and CASPIANET appear to have over-segmentations (see red arrow in Figure 7 (b) and Figure 7 (c)).

The white arrows in Figure 7 (d) and Figure 7 (e) show where the multiscale and multiplanar versions were able to localise the shape of the smaller tumor structure much better than the CASPIANET in Figure 7 (c). Comparing the multiscale with the multiplanar version, adding the multiscale helped with the small tumor structure, but was not able to detect the slight protrusion in the larger tumor structure (see red arrow in Figure 7 (d)). In the combined version, CASPIANET++ (Figure 7 (f)), minor refinements to the outline of the tumor can be seen compared to their individual schemes (see red arrows in Figure 7 (d) and Figure 7 (e)).

For this sample, despite poor contrast between healthy and tumorous tissue, and severely imbalanced tumor classes, CASPIANET++ shows a 10.5% improvement Dice score in the WT region from the baseline DMFNET. Clinically, the improvement presented by CASPIANET++ in demarcating the tumorous region is highly beneficial for treatment planning for radiotherapy or surgical guidance in Gamma Knife surgery, as any over-segmentation will lead to the removal of healthy tissue that could adversely affect brain function.

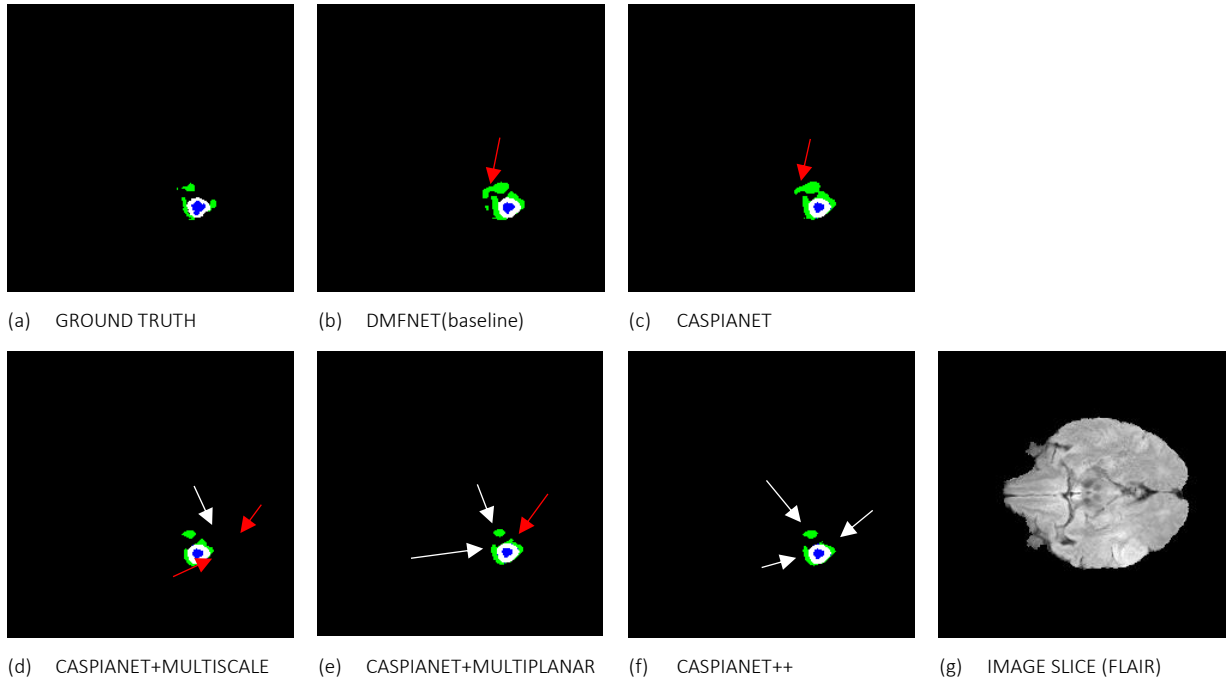

(a) GROUND TRUTH  (b) DMFNET(baseline)  (c) CASPIANET

(d) CASPIANET+MULTISCALE  (e) CASPIANET+MULTIPLANAR  (f) CASPIANET++  (g) IMAGE SLICE (FLAIR)

*Figure 7 The series of images above depict the (a) ground truth, (b) predictions for baseline DMFNET model and (c-f) predictions for CASPIAN variants (g) FLAIR image for Brats18_CBICA_AAL_1 (Slice 65). The Dice scores for ET, WT, TC are 85.003, 77.843, 91.413 (DMFNET) and 84.001, 88.354, 91.825 (CASPIANET++). White arrows denote regions of improved segmentation in each scheme, red arrows denote regions which did not perform well (Blue - Label 1, Green – Label 2, White – Label 4)*

### Visualization of heatmaps of the CASPIANET++ architecture

The heatmaps are generated from a selected feature map from the output of the last CASPIAN block of the CASPIANET++ architecture. As the feature map (120x120) is not spatially equal to that of the input image, a simple up-sampling is performed to produce the heatmaps shown in Figure 8. Regions of high heat intensity (depicted in red) correspond to regions of saliency. In general, heatmaps tend to follow the outline of the whole tumor region, with fairly accurate delineation of the boundaries. In the last row, the heatmap appears slightly smaller, due to the presence of small regions of high contrast artifacts in the opposite hemisphere, which the asymmetric mask will register as being similar.

| Patient data/Slice Index | (a) Slice data | (b) Ground truth | (c) Heatmaps | (d) Predictions |
|---|---|---|---|---|
| Brats18_TCIA08_234_1/ Slice 110 | | | | |
| Brats18_TCIA08_234_1 / Slice 86 | | | | |

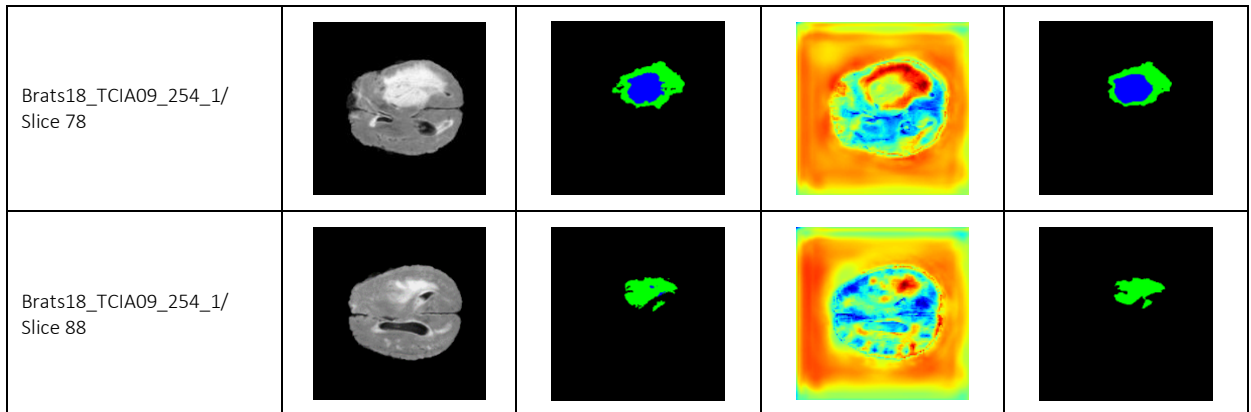

*Figure 8: Image, Ground truth, heatmap and predictions from selected slices (indicated by slice index) obtained from patients Brats18_TCIA08_234_1 and Brats18_TCIA09_254_1 from the BraTS 2018 training set; The heatmaps in (c) are obtained from a selected feature map from the last CASPIAN block of the CASPIANET++, and the predictions in (d) are based on CASPIANET++ (Blue - Label 1, Green – Label 2, White – Label 4)*

Comparison with other brain tumor segmentation methods

In Table II we compare our validation results with the published results from the top 3 architectures from the BraTS 2018 competition. With the exception of CASPIANET++, all performance numbers were obtained from [30]. A PyTorch library available on github (https://github.com/sovrasov/flops-counter.pytorch) was used to obtain the performance values of the CASPIANET++ model. The library was validated with the reported numbers in the DMFNET paper before it was used to report on CASPIANET++. The winner of BraTS 2018, NVDLMED proposed an auxiliary VAE model to reconstruct the encoded image to regularize the main network [21]. The performance of this network is superior in ET regions. One possible factor is the spatial context provided by the large patch size (160x192x168) used during training compared to the other four architectures, DMFNET, No New-Net, CASPIANET++ (128x128x128) and DeepSCAN (5x192x192).

*Table II Results of BraTS2018 validation set obtained from the online validation tool and from published results.*

| Model | Dice Scores (%) | | | Hausdorff95 | | | Performance | |
| --- | --- | --- | --- | --- | --- | --- | --- | --- |
| | ET | WT | TC | ET | WT | TC | Number of floating-point operations (x$10^9$) | Number of parameters (x$10^6$) |
| NVDLMED [21] | **81.73** | 90.68 | 86.02 | 3.8241 | 4.4117 | 6.8413 | 1495.53 | 40.06 |
| No New-Net [20] | 80.66 | 90.92 | 85.22 | 2.74 | 5.83 | 7.20 | 202.25 | 10.36 |
| DeepSCAN [45] | 79.60 | 90.30 | 84.70 | 3.55 | **4.17** | 4.93 | - | - |
| DMFNet [30] | 80.12 | 90.62 | 84.54 | 3.06 | 4.66 | 6.44 | 27.04 | 3.88 |
| CASPIANET++ | 81.08 | **91.20** | **87.60** | **2.66** | 5.05 | **4.91** | 27.22 | 4.23 |

The authors of the baseline DMFNET trained their model with a batch size of 12 on 4 parallel Nvidia GeForce 1080Ti GPUs for 500 epochs [30]. We have achieved better scores using CASPIANET++ with a 50% reduction in the number of epochs and a 2/3 reduction in batch size on 2 parallel NVIDIA GeForce RTX 2080 Ti. The results in Table II show that CASPIANET++ registers an increase in Dice Scores for all 3 regions over DMFNet. Our model appears to converge at a lower number of epochs, attributed by the CASPIAN/CASPIAN++ blocks being driven by CAAM and SAAM, which are a-priori cosine similarity calculations that do not require additional training.

The CASPIANET++ architecture introduces an additional 0.18G of floating-point operations (FLOP) and 0.35M parameters compared to DMFNET. NVDLMED, with its main and auxiliary branch, utilizes the most FLOP and parameters compared to other schemes. In contrast, CASPIANET++ uses 98% less FLOPs and 89% less parameters, while achieving modest improvements in Dice Scores for WT and TC regions and Hausdorff distance for ET and TC regions over NVDLMED.

To determine the statistical significance of these improvements, we use a Wilcoxon paired sample test on the results of the baseline model DMFNET and CASPIANET++. The test results show there is a statistical significance ($p<0.05$) in the improvement of Dice scores in all 3 regions (ET, $p < 0.0001$; WT, $p = 0.0010$; TC, $p = 0.0013$).

### Noisy Student Curriculum Learning

In Table III, we compare the performance of Noisy Student Training [18] with our Noisy Student Curriculum Learning. CASPIANET++ showed marked improvements in the WT and TC regions, but the ET regions registered a 0.65% decrease in Dice Scores compared to NVDLMED. Noisy Student Curriculum Learning was proposed to boost the numbers in this region. The CASPIANET++ model reported in Table II was identified as the teacher model to generate pseudo labels for the unlabeled dataset. Note that all experiments were conducted on the various training schemes using the same teacher model.

*Table III Comparison of different self-training methods using CASPIANET++ as the Teacher and Student model. Unless indicated otherwise, each iteration is trained for a total of 300 epochs.*

| Model | Teacher-Student Training Iterations | Dice Scores (%) | | | Sensitivity (%) | | | Specificity (%) | | | Hausdorff95 | | |
|---|---|---|---|---|---|---|---|---|---|---|---|---|---|
| | | ET | WT | TC | ET | WT | TC | ET | WT | TC | ET | WT | TC |
| NVDLMED [21] (Without Self Training) | - | **81.73** | 90.68 | 86.02 | - | - | - | - | - | - | 3.82 | 4.41 | 6.84 |
| CASPIANET++ (Teacher Model) | - | 81.08 | **91.20** | **87.60** | 82.42 | 92.51 | 86.59 | **99.81** | 99.43 | **99.83** | **2.66** | 5.05 | 4.91 |
| Noisy Student Training | 2 | 81.31 | 90.09 | 86.09 | 83.58 | **93.31** | 85.35 | 99.80 | 99.19 | 99.79 | 2.84 | 7.26 | 4.84 |
| Noisy Student Curriculum Training | 3 | 81.56 | 91.12 | 87.34 | **84.08** | 92.54 | **87.11** | 99.79 | **99.43** | 99.81 | 3.09 | **3.89** | **4.80** |

Table III shows that curriculum learning improves the ET region Dice scores of its teacher model from 81.08% to 81.56% (a 0.48% increase) after 3 iterations. Also, we managed to narrow the gap between the ET scores of NVDLMED to 0.17. In contrast, the Noisy Student Training at iteration 2 exhibits a 0.28% increase in the ET region, but experiences over a 1% decrease in the other two regions, compared to the teacher model. In addition, the noisy student training was erratic and could not predict any tumor regions in certain samples. This was not observed when curriculum learning was used. In comparing the Noisy Student with its curriculum learning counterpart, the specificity essentially remained the same, but the sensitivity increased for the ET and TC regions, indicating that the Noisy Student Curriculum method improved the quality of segmentation.

Noisy Student experiments were conducted by training one iteration of the student model for 300 epochs. The student model's validation result was checked every 10 epochs to select the best student model to become the teacher model in the next iteration. In the Noisy Student Training, we stopped at 2 training iterations as the scores for the WT and TC regions had degraded. This shows that in semantic segmentation problems, it may not be advantageous to introduce noise so early in the self-training iteration cycle. Experiments using stochastic depth to simulate noise in the model did not perform so well in CASPIANET++, possibly because the model is not deep enough to reap the benefits of dropping layers for regularization. Also, adding layers to increase teacher model capacity as suggested in [18] did not yield any benefits while increasing model complexity and training time.

### Generalization ability of CASPIANET++ on the BraTS2020 validation dataset

Table IV presents the validation results obtained from the Noisy Student Curriculum Learning model to predict the BraTS2020 validation set. There was no further fine tuning to our models to obtain these results. The Noisy Student Curriculum model performed well, surpassing the TC region's Dice Scores from the Modality-Pairing method [22], which had placed second in BraTS2020. Compared with CASPIANET++ that was trained solely on BraTS2018 data, training with incrementally-noised data in the Noisy Student Curriculum Learning method enabled the model to generalize to unseen data. In contrast, the Noisy Student Training, which was trained with noise from the first iteration, did not perform as well.

*Table IV Performance results for the Brats2020 validation dataset from the online evaluation platform*

|  | Dice Scores (%) | | | Sensitivity (%) | | | Specificity (%) | | | Hausdorff95 | | |
| --- | --- | --- | --- | --- | --- | --- | --- | --- | --- | --- | --- | --- |
|  | ET | WT | TC | ET | WT | TC | ET | WT | TC | ET | WT | TC |
| Modality-Pairing | **78.50** | **90.70** | 83.70 | 78.30 | 90.10 | 80.40 | **100.00** | **99.90** | **100.00** | 32.25 | **4.39** | 8.34 |
| CASPIANET++ | 77.37 | 89.26 | 81.56 | 77.11 | 89.19 | 79.53 | 99.97 | 99.82 | 99.92 | 27.13 | 7.22 | 9.45 |
| Noisy Student Training | 77.29 | 87.84 | 83.55 | 76.86 | **91.90** | 82.53 | 99.97 | 99.86 | 99.96 | 29.80 | 14.37 | 8.16 |
| Noisy Student Curriculum training | 78.49 | 90.62 | **84.26** | **78.76** | 91.58 | **84.08** | 99.97 | 99.90 | 99.95 | **23.62** | 6.06 | **5.20** |

## CONCLUSION

In this article, we proposed a novel multiscale and multiplanar architecture called CASPIANET++ for accurate segmentation of brain tumors (81.08%, 91.20%, 87.60% respectively for ET, WT and TC), resulting in faster convergence, smaller batch sizes and less GPU resources and marked improvement over the base architecture [30]. This method is different from conventional attention which relies on stochastic methods. Instead, we used the inherent symmetry of the brain to distinguish anomalous regions. With multiscale information from their higher-level counterparts, the attention mechanism is provided with higher scale resolution, which may be lost when the encoder converts spatial to contextual information. To account for tumor non-homogeneity, multiplane information is included as spatial asymmetry from a single axis does not provide a holistic picture of the tumor regions, particularly at boundary regions. The result is a lightweight attention block which combines a-priori information with excitation to amplify strong regions of dissimilarity and suppresses non-pertinent information.

ET regions have always registered the lowest Dice scores. In efforts to improve the Dice score in the ET region, we used Noisy Student training. This method was first used in classification of everyday objects – but we ran experiments to see how it would fare in a multiclass brain tumor segmentation problem in a low data regime. The Noisy Student improved the ET scores at the expense of lower scores in the WT and TC regions. This suggests that bombarding the network with noise early in the self-training iterations destabilizes the network, preventing it from generalizing. Also, the drop in the WT and TC scores in the first iteration may be symptomatic of catastrophic forgetting. In the original Noisy student method [18], which led to better ImageNET classification, strong emphasis was placed on using as much noise as possible from the first iteration of the student model. For brain tumor segmentation tasks, the host of noising techniques applied at the onset may have led the student model to "forget" the knowledge from the relatively noise-free teacher model. Hence, a new curriculum learning paradigm was introduced to Noisy Student learning by incrementally adding noise to the training samples, leading to better scores in the ET region (81.56%, 91.12%, 87.34% respectively for ET, WT and TC). This observation is reiterated by the BraTS2020 validation performance. Our results from the Noisy Student Curriculum learning paradigm suggest that incrementally introducing noise to iteratively train a segmentation network with scarce data may lead to gains in accuracy for other tumor segmentation problems.

Our results therefore suggest the CASPIANET++ architecture and the Noisy Student Curriculum Learning can be applied independently or combined to form a two-prong approach to segment tumor regions in medical images in the presence of small annotated and unannotated datasets. As we have demonstrated, the CASPIAN++ layer can be incorporated into any architecture to improve accuracy with minimal impact to computational requirements. The results can be further enhanced by applying a Noisy Student Curriculum Learning to unlabeled data. Here, the flexibility of the "curriculum" allows the method to be adapted to the problem. In our future work, we plan to determine how different definitions of "curriculum" can be used to train the student model to further improve the accuracy.

**Compliance with ethical standards**

**Conflict of interest** None declared

**Acknowledgements**

This study was funded by Sunway Medical Centre, under grant ENG/SUNMED/11-2019/008.